\documentclass[twocolumn]{aastex63}
\usepackage{apjfonts} 
\usepackage{natbib}
\usepackage{graphicx}
\usepackage{amsmath,amsfonts}
\usepackage{multirow}
\usepackage{booktabs}
\usepackage{sidecap}
\usepackage{color}
\usepackage{xspace}
\usepackage{tipa} 

\usepackage{hyperref}

\usepackage[toc,page]{appendix}

\usepackage{float}

\graphicspath{{}}

\newcommand{\change}[1]{\textcolor{black}{#1}}

\newcommand{\Limacon}{Lima\c{c}on\xspace}
\newcommand{\limacon}{lima\c{c}on\xspace}
\newcommand{\LIMACON}{LIMA\c{C}ON\xspace}
\newcommand{\limacons}{lima\c{c}ons\xspace}
\newcommand{\inclination}{\theta_\textrm{o}}

\setcitestyle{notesep={; }}

\def\sgra{Sgr~A$^{\ast}$}

\def\lsim{\mathrel{\raise.3ex\hbox{$<$\kern-.75em\lower1ex\hbox{$\sim$}}}}
\def\gsim{\mathrel{\raise.3ex\hbox{$>$\kern-.75em\lower1ex\hbox{$\sim$}}}}
\def\gtwid{\mathrel{\raise.3ex\hbox{$>$\kern-.75em\lower1ex\hbox{$\sim$}}}}
\def\proptwid{\mathrel{\raise.3ex\hbox{$\propto$\kern-.75em\lower1ex\hbox{$\sim$}}}}

\defcitealias{EHT1}{EHT1}
\defcitealias{EHT2}{EHT2}
\defcitealias{EHT3}{EHT3}
\defcitealias{EHT4}{EHT4}
\defcitealias{EHT5}{EHT5}
\defcitealias{EHT6}{EHT6}

\newcommand{\bhi}{Black Hole Initiative, Harvard University, 20 Garden Street, Cambridge, MA 02138, USA}
\newcommand{\cfa}{Center for Astrophysics $\vert$ Harvard-Smithsonian, 60 Garden Street, Cambridge, MA 02138, USA}
\newcommand{\umb}{University of Massachusetts Boston, 100 William T.~Morrissey Blvd, Boston, MA 02125, USA}

\begin{document}

\title{On the approximation of the black hole shadow with a simple polar curve} 
\shorttitle{On the approximation of the black hole shadow with a simple polar curve}

\shortauthors{Joseph R. Farah et al.}
\correspondingauthor{Joseph R. Farah}
\email{joseph.farah@cfa.harvard.edu}

\author[0000-0003-4914-5625]{Joseph R. Farah}
\affiliation{\cfa}
\affiliation{\bhi}
\affiliation{\umb}

\author[0000-0002-5278-9221]{Dominic W. Pesce}
\affiliation{\cfa}
\affiliation{\bhi}

\author[0000-0002-4120-3029]{Michael D. Johnson}
\affiliation{\cfa}
\affiliation{\bhi}

\author[0000-0002-9030-642X]{Lindy Blackburn}
\affiliation{\cfa}
\affiliation{\bhi}




\begin{abstract}
\noindent A black hole embedded within a bright, optically thin emitting region imprints a nearly circular ``shadow'' on its image, corresponding to the observer's line-of-sight into the black hole. The shadow boundary depends on the black hole's mass and spin, providing an observable signature of both properties via high resolution images. However, \change{standard expressions for the shadow boundary are most naturally parametrized by Boyer-Lindquist radii rather than by image coordinates}. We explore simple, approximate parametrizations for the shadow boundary using ellipses and a family of curves known as \textit{\limacon{s}}. We demonstrate that these curves provide excellent and efficient approximations for all black hole spins and inclinations. In particular, we show that the two parameters of the \limacon naturally account for the three primary shadow deformations resulting from mass and spin: size, displacement, and asymmetry. These curves are convenient for \change{parametric model fitting directly to interferometric} data, they reveal the degeneracies expected when estimating black hole properties from images \change{with practical measurement limitations}, and they provide a natural framework for parametric tests of the Kerr metric using black hole images.
\end{abstract}

\keywords{ black hole physics --- radio continuum: Galaxy: nucleus --- techniques: interferometric}


\section{Introduction}
\label{sec:introduction}

When surrounded by optically thin emitting material, a black hole produces a nearly circular ``shadow'' on the image seen by a distant observer, corresponding to the observer's line-of-sight into the black hole \citep[see, e.g.,][]{bardeen_1973,Jaroszynski_1997,deVries_2000,Falcke_2000,Perlick_2004, Cunha_2018,Narayan_2019}. 
The size and shape of the shadow encode information about the black hole's angular momentum, inclination relative to the observer, and mass. Moreover, the shadow boundary is achromatic and independent of the emission and astrophysical details of a source. 
However, the angular diameter of the shadow is only ${\approx} 10 G M/(c^2 D) \approx 10\,\mu{\rm as} \left(\frac{M}{10^9M_\odot} \right) \left( \frac{D}{10\,{\rm Mpc}} \right)^{-1}$ in size, where $G$ is the gravitational constant, $M$ is the mass of the black hole, $c$ is the speed of light, and $D$ is the distance from the observer \citep{Hilbert1917a,Synge_1966, bardeen_1973, Takahashi_2004}. Thus, even for nearby supermassive black hole \change{candidates}, very high angular resolution is required to study the shadow. 

\begin{figure*}[htb!]
\centering
\includegraphics[width=\textwidth]{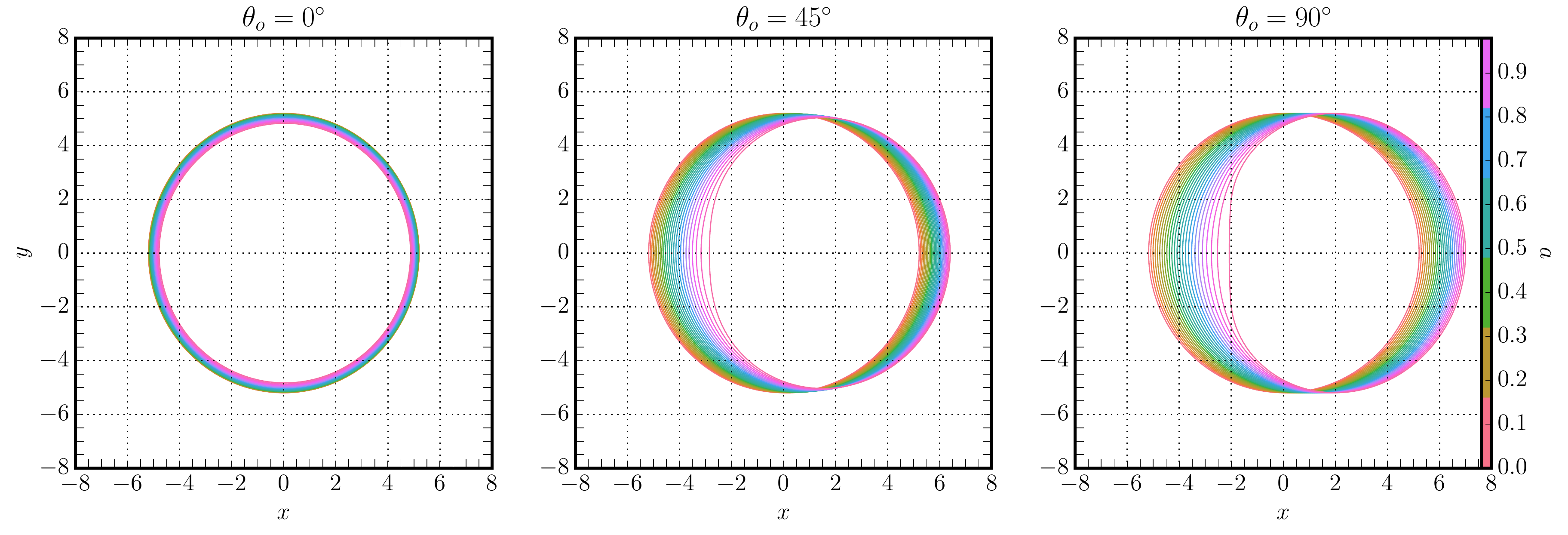}
\caption{Black hole shadow boundary as a function of spin and inclination. The dimensionless spin magnitude $a$ is sampled uniformly from 0 to 1 in steps of 0.02, while the three panels correspond to inclinations $\inclination=0^\circ$ (left), $\inclination=45^\circ$ (center), and $\inclination=90^\circ$ (right).} 
\label{fig:shadows}
\end{figure*}

This resolution was recently achieved by the Event Horizon Telescope, using Very Long Baseline Interferometry (VLBI) at 1.3mm wavelength to image the supermassive black hole in the galaxy M87 (\citealt{EHT1,EHT2,EHT3,EHT4,EHT5,EHT6}; hereafter EHT1-6). The EHT observations revealed a bright, asymmetric ring, corresponding to a black hole with $M = (6.5 \pm 0.7) \times 10^9\,M_\odot$ at a distance $D = (16.8 \pm 0.8)\,{\rm Mpc}$. 
While the EHT does not yet provide significant constraints on the shadow shape or asymmetry, planned extensions to shorter wavelengths and longer baselines will achieve even finer resolution and a greater reduction of the astrophysical foreground. Thus, it is essential to develop tools and techniques that are capable of more detailed model fitting and inference and to assess what can be learned about a black hole from higher resolution observations of its shadow.

To this end, many authors have quantified the shadow of a Kerr black hole using simple diagnostics such as mean diameter, asymmetry, and displacement from the origin \citep[e.g.,][]{Takahashi_2004,Johannsen_2010,Chan_2013, Grenzebach_2015, Tsupko_2017}.  Other authors have moved beyond these diagnostics, presenting general approximation frameworks to model arbitrary shadows using a Legendre expansion \citep{Abdujabbarov_2015} or a basis of principal components \citep{Medeiros_2019}. Several authors have explored simpler parametric approximations for the shadow, either using approximations for the full relativistic calculation of the shadow \citep{Cunha_2018} or using families of simple curves \citep{vries_2003}. Here, we expand this latter approach, developing approximations for the shadow using simple polar curves. We show that these curves provide excellent representations for the exact shadow shape, while offering a variety of concrete advantages. Relative to simple shadow diagnostics, these curves have the advantage of completely characterizing the information in the shadow, thereby highlighting potential parameter degeneracies that are expected to persist even for high resolution images. They also provide tools for model fitting directly to interferometric data, and they can be used as a parametric framework to assess the validity of the Kerr shadow assumption.

We begin, in \autoref{sec:kerr_construction}, with a brief introduction to the shadow and its properties. Next, we explore approximations to the shadow using a shifted ellipse (\autoref{sec:ellipse}), a \limacon (\autoref{sec:limacon}), and the convex hull of a shifted \limacon (\autoref{sec:convex_hull}). In \autoref{sec:discussion}, we compare these results and discuss their implications for the inference of black hole parameters with high resolution images. We summarize our results in \autoref{sec:summary}.

\section{The Shadow of a Kerr Black Hole}
\label{sec:kerr_construction}

The Kerr metric \citep{kerr_1963} describes the vacuum spacetime around a rotating, non-charged black hole of mass $M$ and dimensionless spin $a = J c/M^2$, where $J$ is the angular momentum of the black hole and $0 \leq a \leq 1$. A crucial feature of this metric is that photons near the black hole can orbit on spheres with constant Boyer-Lindquist radius, $r$. For a Schwarzschild black hole, all spherical photon orbits have $r=3 GM/c^2$, while a spinning black hole has $r_{+} \leq r \leq r_{-}$, where the radii of the prograde ($+$) and retrograde ($-$)  equatorial photon orbits are
\begin{equation}
	r_{\pm} = 2\frac{G M}{c^2} \left \{ 1+\cos \left[\frac{2}{3}\arccos\left(\mp |a|\right)\right] \right \}.
    \label{eq:rph}
\end{equation}
For simplicity, we will set $G = c = M = 1$ for the remainder of our discussion and will express later quantities in units of $r_{\rm g} \equiv G M/c^2$.

Null geodesics in the Kerr metric are fully characterized by two constants of motion: the energy-rescaled angular momentum about the spin axis $\ell = L/E$, and the energy-rescaled Carter constant $\chi = Q/E^2$. For a spherical photon orbit at radius $r$, these are given by \citep[see, e.g.,][]{bardeen_1973,Teo_2003,Cunha_2018}
\begin{align}
          \ell &= -\frac{ r^3 + a^2 r + a^2 - 3 r^2}{a(r-1)},\\
\nonumber \chi &= \frac{r^2}{r^2-a^2} \left( 3r^2 + a^2 - \ell^2 \right).
\end{align}

A slight perturbation from a spherical orbit will exponentially diverge away from the photon orbit. Slightly tighter orbits plunge into the black hole, while slightly larger orbits escape to infinity. This instability produces a brightness enhancement at the boundary of the shadow if the region $r_{+} \leq r \leq r_{-}$ is populated by emitting material and is optically thin \citep[see, e.g.,][]{Gralla_2019,Gralla_Lupsasca_2019,Johnson_2020}. A distant observer viewing the black hole at inclination $\inclination$ relative to the spin axis will then see this asymptotic brightness enhancement at Cartesian screen coordinates given by \citet{bardeen_1973}: 
\begin{align}
\label{eq:alpha_beta} 
x(r) &= -\ell/\sin\inclination\\
\nonumber &= \frac{r^3 + a^2 r + a^2 - 3 r^2}{a(r-1) \sin \inclination},\\
\nonumber y(r) &= \pm \sqrt{ \chi + a^2 \cos^2\inclination - \ell^2/\tan^2\inclination} \\
\nonumber &= \pm \frac{1}{a(r-1)}\Big\{ a^4 (r-1)^2 \cos^2 \inclination -[a^2 (r+1) \\
\nonumber &\qquad {} + (r-3)r^2]^2 \cot^2 \inclination - r^3 [r(r-3)^2 -4a^2]\Big\}^{1/2}. 
\end{align}
Here, the projected spin direction on the observer's screen lies along the $y$-direction. The interior of this curve corresponds to the observer's line of sight into the black hole. 

Figure~\ref{fig:shadows} shows example shadows at three inclinations and all spins. For a Schwarzschild black hole ($a=0$), the shadow is circular and has a radius on the image of $\sqrt{x(r)^2 + y(r)^2} = \sqrt{27} r_{\rm g}$ \citep{Hilbert1917a,Synge_1966}. As $a$ increases, the shadow is displaced horizontally and flattens on the side approaching the origin. This effect becomes most pronounced in the limit $a\to1$ and $\inclination\to\pi/2$ (a maximally spinning black hole viewed edge-on).  

While \autoref{eq:alpha_beta} defines the exact boundary of the black hole shadow, it is inconveniently parametrized by the corresponding (Boyer-Lindquist) \emph{emission radius} rather than by a natural image coordinate such as the polar angle. Next, we will explore simple polar approximations for the shadow.

\section{Ellipse approximation}
\label{sec:ellipse}

\subsection{Definitions}
\label{sec:ellipse_motivation}

Our first polar curve to describe the black hole shadow is an ellipse, including a displacement parameter. In addition to its simple parametric form, the ellipse has convenient Fourier relationships to a circle, making it particularly effective as a tool for model fitting directly to interferometric data. Specifically, by the Fourier scaling theorem, any circular image with a known Fourier transform can be stretched by a factor $f$ to create an ellipse, with a Fourier transform given by stretching baselines and rescaling visibility amplitudes of the circular image by the factor $f^{-1}$ along the same direction. For instance, any circular ring model used to fit the EHT observations in \citetalias{EHT6} could be trivially adapted to allow elliptical shapes \citep[see also][]{Kamruddin_2013,Benkevitch_2016}. 

We parametrize the ellipse as $\left( x(\tau), y(\tau) \right)$, where
\begin{align} 
    x(\tau)   &= r_0(\tau)\cos(\tau) + \delta,\\
    y(\tau)   &= r_0(\tau)\sin(\tau),\\
    r_0(\tau) &= \frac{r_\parallel r_\perp}{\sqrt{r_\parallel^{2} \cos^2\tau+r_\perp^{2}\sin^2\tau}},
\end{align}
Here, $\delta$ is a shift in the positive $x$-direction, $r_\parallel$ is the ellipse radius along the spin axis, $r_{\perp}$ is the ellipse radius orthogonal to the spin axis, and $\tau \in [0,2\pi]$ parametrizes the curve. If $\delta =0$, then $\tau$ is equal to the polar angle $\varphi$.

\subsection{Fidelity of Shadow Fits}
\label{sec:ellipse_fidelity}

For every shadow given by a pair of black hole parameters $a$ and $\inclination$, we determined the best-fitting ellipse parameters by minimizing the root-mean-squared (RMS) radial residual, weighted by arc length ${\rm d}\ell(\varphi) = \sqrt{ r(\varphi)^2 + r'(\varphi)^2 } {\rm d}\varphi$. Namely, we minimized
\begin{align}
    \sigma\left(r_\perp, r_{\parallel}, \delta\right) &\equiv \sqrt{ \frac{\int_0^{2\pi} {\rm d}\varphi\left[ \Delta x(\varphi)^2 + \Delta y(\varphi)^2 \right] \sqrt{ r(\varphi)^2 + r'(\varphi)^2 } }{\int_0^{2\pi} {\rm d}\varphi \sqrt{ r(\varphi)^2 + r'(\varphi)^2 } }}.
\end{align}
Weighting by arc length gives results that are independent of how the curve is parametrized, including the assumed centroid. Other weighting choices, such as ${\rm d}\tau$, ${\rm d}\varphi$, or the line element ${\rm d}r$ resulting from parametrization by Boyer-Lindquist radius (\autoref{eq:alpha_beta}) can heavily upweight particular segments of the shadow curve.

\begin{figure}[]
\includegraphics[width=\columnwidth]{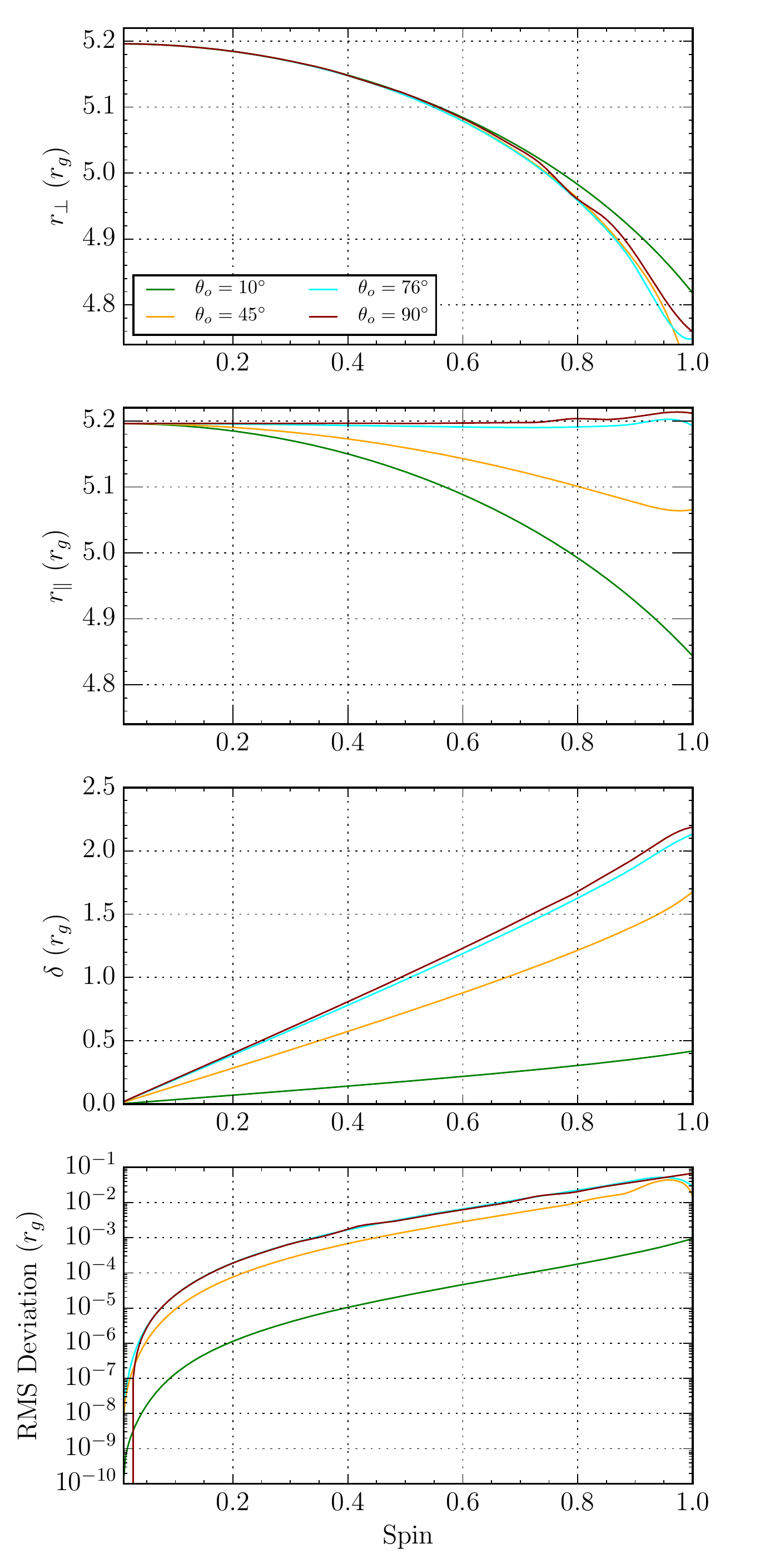}
\caption{Ellipse parameters that best fit the shadow boundary as a function of black hole spin and inclination. From top to bottom, panels show the ellipse radius perpendicular to the projected spin axis ($r_{\perp}$), the ellipse radius parallel to the projected spin axis ($r_{\parallel}$), the ellipse displacement ($\delta$), and the RMS error of the fitted curve weighted by arc length ($\sigma = \sqrt{\langle \delta r(\varphi)^2 \rangle_\ell}$). }  
\label{fig:ellipse_tower_plot}
\end{figure}

\autoref{fig:ellipse_tower_plot} shows the best-fitting ellipse parameters and RMS radial residual at four inclinations and all spins. In general, the parameters vary smoothly and monotonically with spin except at very high spin. Note that the shadow radius orthogonal to the spin axis monotonically decreases with spin, while the shadow axis parallel to the spin axis decreases at low inclination but increases at high inclination.

At all non-zero inclinations, the RMS residual increases with spin. As shown in \autoref{fig:shadows}, the left shadow edge flattens as spin increases, which the ellipse cannot reproduce. Nevertheless, the RMS residual is $\sigma \lsim 0.1 r_{\rm g}$ for all cases, corresponding to a fractional radial error of approximately $2\%$ or less. For a black hole viewed at low inclination or a black hole with low spin, the accuracy of the ellipse is excellent. For the supermassive black hole in M87, which has $\inclination \approx 17^\circ$ \citep{Walker_2018} and $r_{\rm g}/D \approx 3.8\,\mu{\rm as}$ \citepalias{EHT6}, the ellipse has $\sigma \lsim 0.005 r_{\rm g}$ ($\sigma/D \lsim 0.02\,\mu{\rm as}$) or a fractional radial error of approximately $0.1\%$.

\section{\LIMACON approximation}
\label{sec:limacon}

\subsection{Definitions}
\label{sec:limacon_motivation}

Our second polar curve to describe the black hole shadow is a \limacon\footnote{Pronounced \textipa{/" lIm@s6n/}}, formed by tracing a fixed point on a circle as it rotates around another circle of equal radius displaced by one diameter. The \limacon has a simple polar representation that depends on only two parameters:
\begin{align}
\label{eq:limacon} 
    r(\varphi) \equiv \lambda_1 (1 + \lambda_2\cos\varphi). 
\end{align}
Here, $\varphi$ is the polar angle, $\lambda_1 > 0$ is a scale parameter, while $\lambda_2$ contributes both asymmetry and displacement. 
Note that the \limacon is convex when $|\lambda_2| \leq 1/2$. For $|\lambda_2| > 1$, the \limacon intersects itself at the origin for $\varphi = \pm \cos^{-1}\left(-\lambda_2^{-1}\right)$. For $\lambda_2=1$, the \limacon is exactly a cardioid, with a cusp at $\varphi=\pi$. \change{Because changing the sign of $\lambda_2$ simply gives an image reflection, we will assume that $\lambda_2 > 0$ for the remainder of this paper.} 

\begin{figure}[t]
\includegraphics[width=\columnwidth]{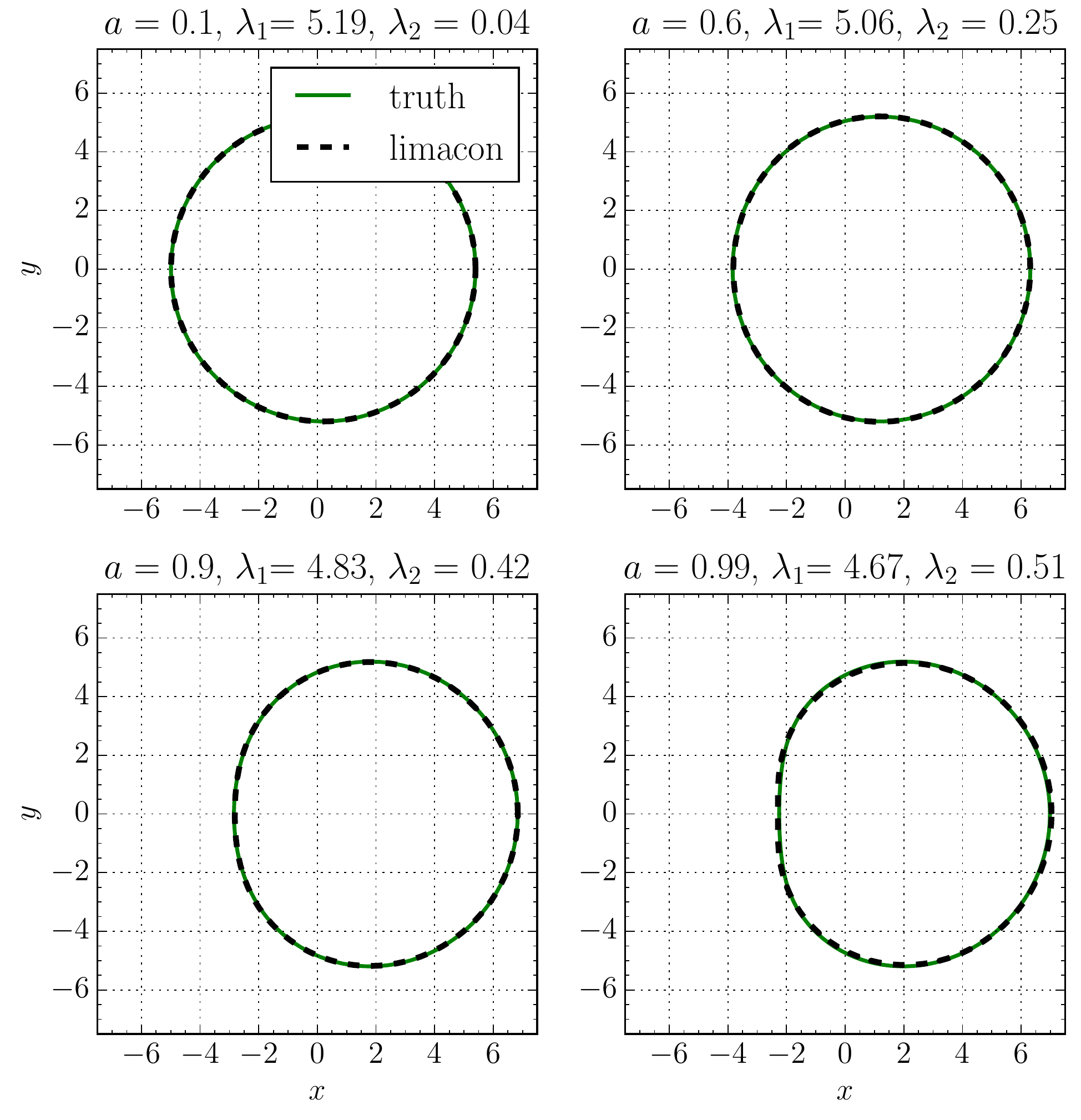}
\caption{Comparison between exact shadows (green, solid) and their best-fitting \limacons (black, dashed) for $\varphi=90^\circ$ (i.e., edge-on viewing). The \limacon parameter $\lambda_2$ increases with spin, producing both the asymmetry and centroid displacement seen in the exact shadow.
}
\label{fig:deformation_quantity_calculation}
\end{figure}

The \limacon exhibits similar behavior to the shadow, shifting itself over from the origin by a similar amount and producing an edge on the left side as it progresses further from the origin. The similarity between the shadow and the \limacon is demonstrated in \autoref{fig:deformation_quantity_calculation} (and is quantified in \autoref{sec:limacon_fidelity}). Hence, unlike the ellipse, we do \emph{not} include a shift parameter for fits to the pure \limacon. \cite{vries_2003} proposed the \limacon as an approximation to the shadow of a Kerr-Newman black hole and provided several example fits, demonstrating excellent agreement. The quality of these fits declined with increasing spin. In Section~\ref{sec:convex_hull}, we will generalize the \limacon by including a shift parameter and an additional operation to ensure convexity; these modifications substantially improve the quality of the shadow fits.

\subsection{Characteristics}
\label{sec:limacon_defq}

We now compute some simple derived quantities for the \limacon, to clarify how the parameters $\lambda_1$ and $\lambda_2$ affect the curve and relate to shadow properties. 

The simplest characteristics of the \limacon are quantities weighted uniformly by angle with respect to the origin of the \limacon coordinates. The mean radius is $\langle r \rangle_\varphi = \lambda_1$, while the standard deviation of the radius is $\sqrt{\langle \Delta r^2 \rangle_\varphi} = \lambda_1 \lambda_2/\sqrt{2}$. The horizontal displacement is $\Delta \equiv \frac{x_{\textrm{max}} + x_{\textrm{min}}}{2} = \lambda_1\lambda_2$.

However, for comparisons with other families of curves (including the Kerr shadow), it is convenient to instead compute quantities with respect to coordinates that are centered on the curve, according to the horizontal displacement, and weighted by arc length. For example,
\begin{align}
    \langle r_\Delta \rangle_\ell &\approx \lambda_1 \left( 1 + \frac{\lambda_2^2}{4}  - \frac{\lambda_2^4}{64}  + \mathcal{O}(\lambda_2^6) \right).
\end{align}
\change{It is also useful to evaluate the curve radii along directions orthogonal ($r_{\perp}$) and parallel ($r_{\parallel}$) to the vertical axis (i.e., the projected spin axis). These radii are independent of the assumed centroid or curve weighting and have been discussed extensively for the specific case of a black hole's shadow \citep[e.g.,][]{Grenzebach_2015, Tsupko_2017}. For the \limacon, 
\begin{align}
\label{eq::rparperp}
     r_{\perp} &\equiv \frac{1}{2}\left( {\rm max}_\varphi\left[ r(\varphi) \cos(\varphi) \right] - {\rm min}_\varphi\left[ r(\varphi) \cos(\varphi) \right] \right)\\
\nonumber      &= \lambda_1 \times
     \begin{cases} 
      1 & \lambda_2 \leq 1/2 \\
      \frac{\left(1 + 2\lambda_2\right)^2}{8 \lambda_2} & \lambda_2 > 1/2,
  \end{cases}
\end{align}
and 
\begin{align}
  r_{\parallel} &\equiv \rm{max}_\varphi\left[ r(\varphi) \sin(\varphi) \right]\\
\nonumber         &= \lambda_1 \frac{\left( 3 + \sqrt{1 + 8\lambda_2^2} \right) \sqrt{\sqrt{1 + 8\lambda_2^2} + 4\lambda_2^2 - 1 }}{8\sqrt{2}\lambda_2}\\
\nonumber         &\approx \lambda_1 \left( 1 + \frac{1}{2}\lambda_2^2 - \frac{5}{8} \lambda_2^4 + \dots \right). 
\end{align}}

\subsection{Fidelity of Shadow Fits}
\label{sec:limacon_fidelity}

\begin{figure}[]
\includegraphics[width=\columnwidth]{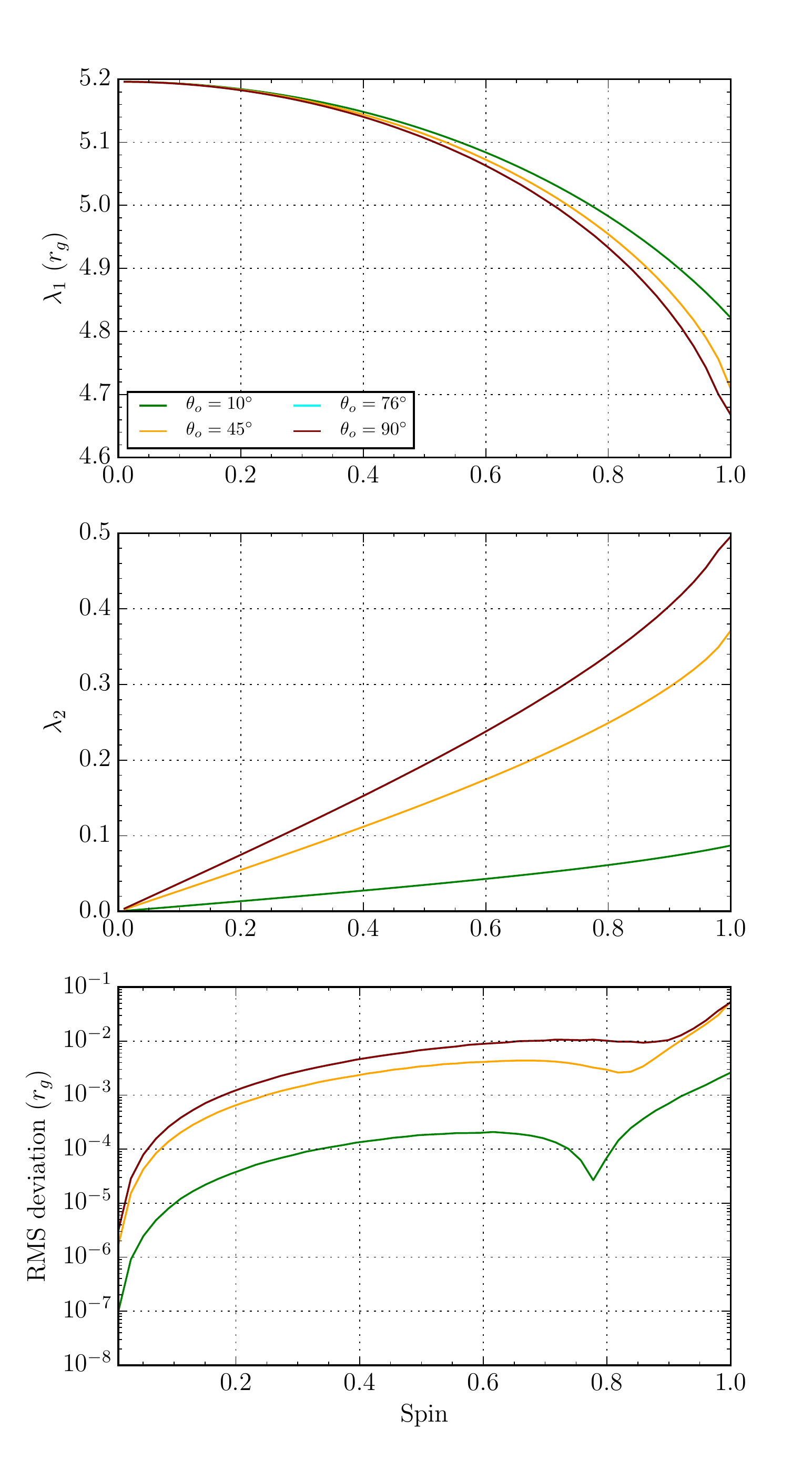}
\caption{
\Limacon parameters that best fit the shadow boundary as a function of black hole spin and inclination. From top to bottom, panels show the \limacon scale parameter $\lambda_1$, the \limacon asymmetry parameter $\lambda_2$, and the RMS error of the fitted curve $\sigma = \sqrt{\langle \delta r(\varphi)^2 \rangle_\ell}$. Unlike the ellipse, both \limacon parameters vary monotonically with spin and inclination (although the RMS error is not monotonic in spin). 
}
\label{fig:limacon_tower_plot}
\end{figure}

\autoref{fig:limacon_tower_plot} shows the best-fitting \limacon parameters at four inclinations and for all spins, obtained by minimizing the RMS radial residual weighted by arc length as described in Section~\ref{sec:ellipse_fidelity}. The \limacon parameters are monotonic in spin and inclination, and the \limacon generally outperforms the ellipse in approximating the shadow despite requiring one fewer parameter. 
For $a < 0.95$, \limacons fit the shadow with $\sigma \lsim 10^{-2} \ r_g$, or a fractional radial error of less than $0.2\%$.  

\begin{figure}[]
\centering
\includegraphics[width=\columnwidth]{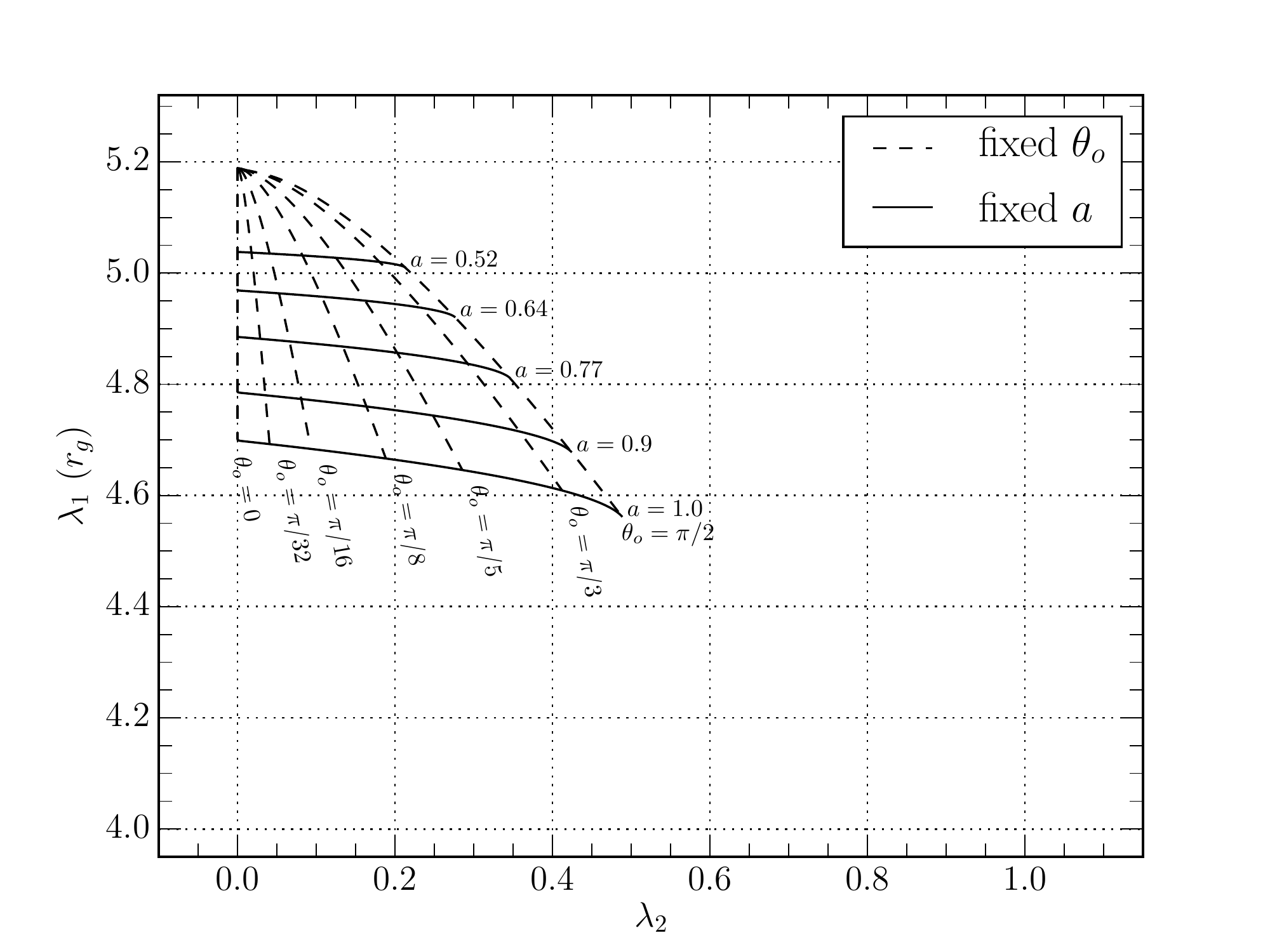}
\caption{Best-fitting \limacon parameters $(\lambda_1, \lambda_2)$ for all spins and inclinations. 
}
\label{fig:altgrav}
\end{figure}

Figure~\ref{fig:altgrav} shows the best-fitting \limacon parameters as a function of black hole spin and inclination. Importantly, the mapping is one-to-one, so the \limacon parameters uniquely determine the black hole spin and inclination if the mass-to-distance ratio of the black hole is known. Because the Kerr metric occupies only a small part of the $(\lambda_1, \lambda_2)$ parameter space, a \limacon fit can also be used to assess the validity of the Kerr assumption for the shadow and to quantify departures from Kerr.

\section{Convex hull \LIMACON approximation}
\label{sec:convex_hull}

\begin{figure}[t]
\centering
\includegraphics[width=\columnwidth]{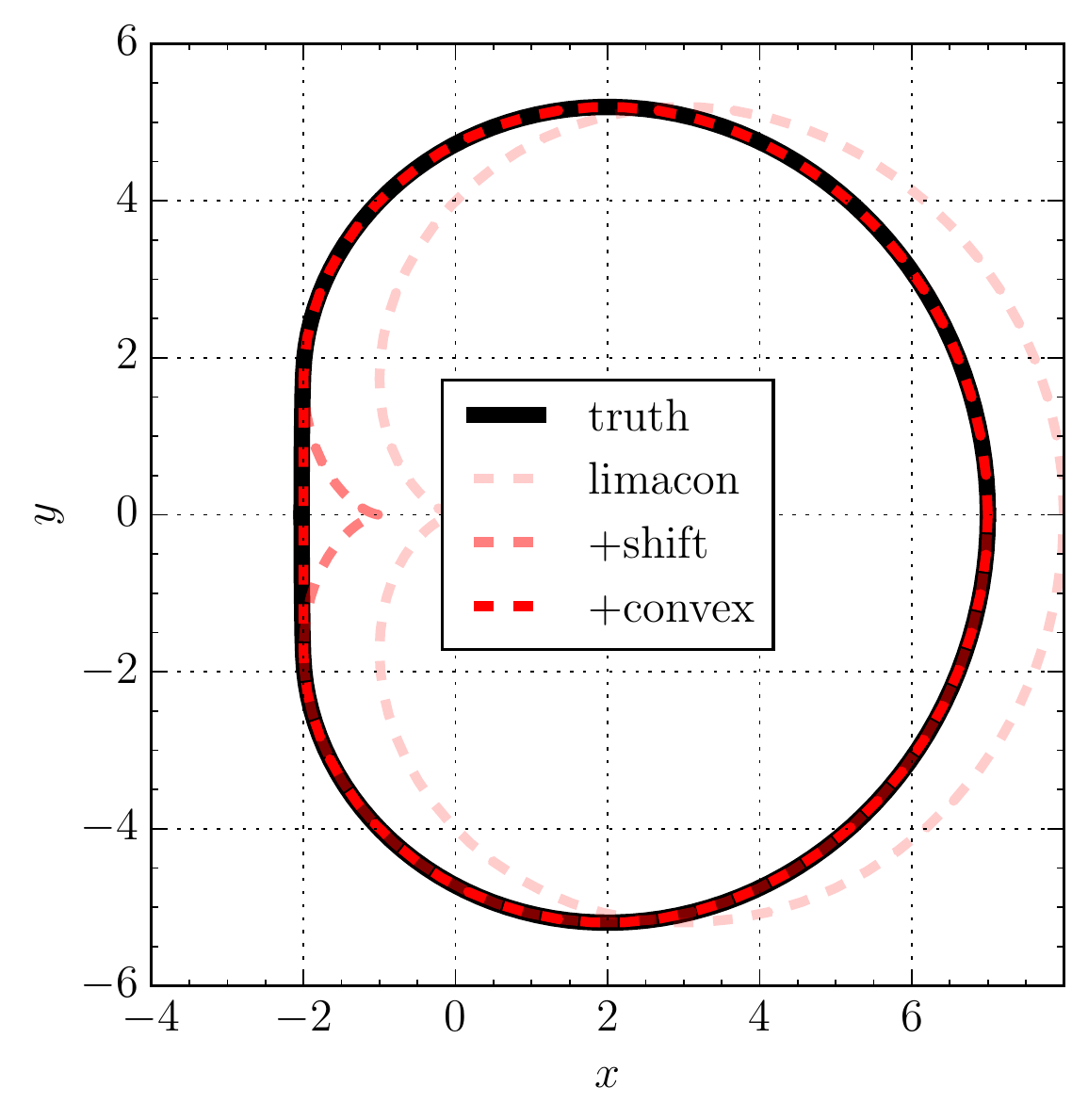}
\caption{In the maximal-spin, edge-on limit, the shadow boundary (black) is the convex hull of a shifted \limacon with $\lambda_1 = 4$, $\lambda_2=1$, and $\delta=-1$. Because $\lambda_2 > 1/2$, the pure \limacon curls into a cardioid. Hence, both the displacement and convex hull are necessary to exactly match the shadow. 
}
\label{fig:cardioid}
\end{figure}

\subsection{Definitions}
\label{sec:convex_motivation}

As shown in Figure~\ref{fig:limacon_tower_plot}, the \limacon provides an excellent approximation to the shadow but is least accurate at high spin and inclinations approaching edge-on. Remarkably, the shape of the shadow in the limit of maximal spin and edge-on inclination is precisely a \limacon with $\lambda_1 = 4 r_{\rm g}$ and $\lambda_2 = 1$, with two modifications. First, it is horizontally displaced from the origin (see Figure~\ref{fig:cardioid}). Namely, it takes the form,
\begin{align}
\label{eq:cos44}
    r(\varphi; \delta = -1) = 4(1 + \cos\varphi),
\end{align}
where $\delta = -1$ indicates that $r$ and $\varphi$ are polar coordinates defined relative to an origin that is horizontally displaced by $\delta = -1$. Second, the shadow for these asymptotic parameters has a hard, flat edge joining the discontinuous portion in the region $\varphi\in[2\pi/3, 4\pi/3]$, which deviates from the signature cusp of the $\lambda_2 = 1$ \limacon. This behavior can be reproduced by taking the convex hull of the shifted \limacon. \change{\autoref{eq::rparperp} then gives that the shadow of an extremal black hole viewed edge-on has $r_{\perp} = \frac{9}{2} r_{\rm g}$ and $r_{\parallel} = 3\sqrt{3} r_{\rm g}$. This result was derived by \citet{Grenzebach_2015}, who provide a general expression for $r_{\perp}$ and $r_{\parallel}$ for a black hole of any spin viewed edge-on. }

Motivated by the exact form of the shadow in this limit, our third family of parametric curves is the convex hull of a shifted \limacon. This curve adds a single parameter, the shift $\delta$; the convex hull is only relevant for $|\lambda_2| > 1/2$, where a flat edge is introduced at $\varphi = \pm \arccos\left(-1/2\lambda_2\right)$ or at $x_\textrm{min} = \delta-\lambda_1/4 \lambda_2$. 
In the Kerr parameter space, the convex hull is only needed for $a\gtrapprox 0.95$.

\subsection{Fidelity of Shadow Fits}
\label{sec:convex_fidelity}

\begin{figure}[t]
\centering
\includegraphics[width=\columnwidth]{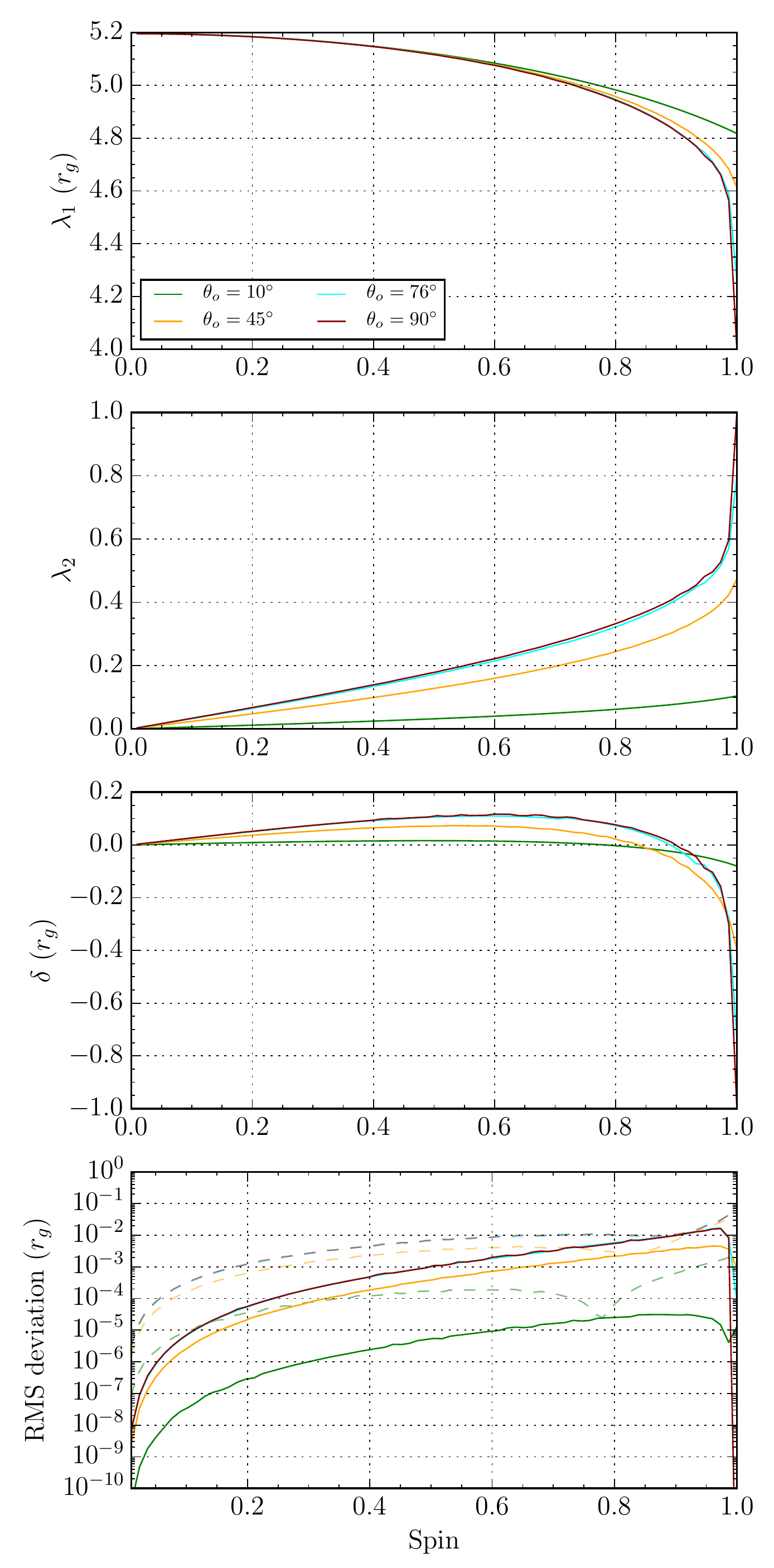}
\caption{
Convex hull \limacon parameters that best fit the shadow boundary as a function of black hole spin and inclination. From top to bottom, panels show the scale parameter $\lambda_1$, the asymmetry parameter $\lambda_2$, the displacement $\delta$, and the RMS radial residual $\sqrt{\langle \delta r(\varphi)^2 \rangle_\ell}$. For comparison, the RMS error of the \limacon is also shown as dashed lines (see Figure~\ref{fig:limacon_tower_plot}). All inclinations and spins have a point near $a=0.8$ where $\delta=0$; at this location the \limacon and convex hull \limacon fit equally well unless $\lambda_2 > 1/2$ because only difference improvement in the convex hull \limacon is its additional shift parameter.
}
\label{fig:convexhull_tower_plot}
\end{figure}

Using the methodology described in \autoref{sec:ellipse_fidelity}, \limacons were fit to shadow shapes at four inclinations for all spin. The results are shown in \autoref{fig:convexhull_tower_plot}.

The convex hull \limacon offers fits that can be an order of magnitude better than those of the unshifted \limacon or the ellipse. Size and asymmetry parameter variation are monotonic, but the curves quantifying the horizontal shift of the approximation increase in concavity while shifting downwards as inclination approaches $\inclination=\pi/2$. All inclinations and spins have a point at around $a=0.8$ where $\delta=0$; here, the best-fitting convex hull \limacons match the normal, unshifted \limacon provided that $\lambda_2 < 1/2$. Indeed, this agreement can be seen in the intersections with the lighter, dotted curves in the bottom panel of \autoref{fig:convexhull_tower_plot} representing the resulting accuracy of the \limacon fits from \autoref{fig:limacon_tower_plot}. The fit fidelity generally worsens as the asymmetry of the true shadow increases, but it improves sharply near maximal spin. 

\begin{figure}[t]
\centering
\includegraphics[width=\columnwidth]{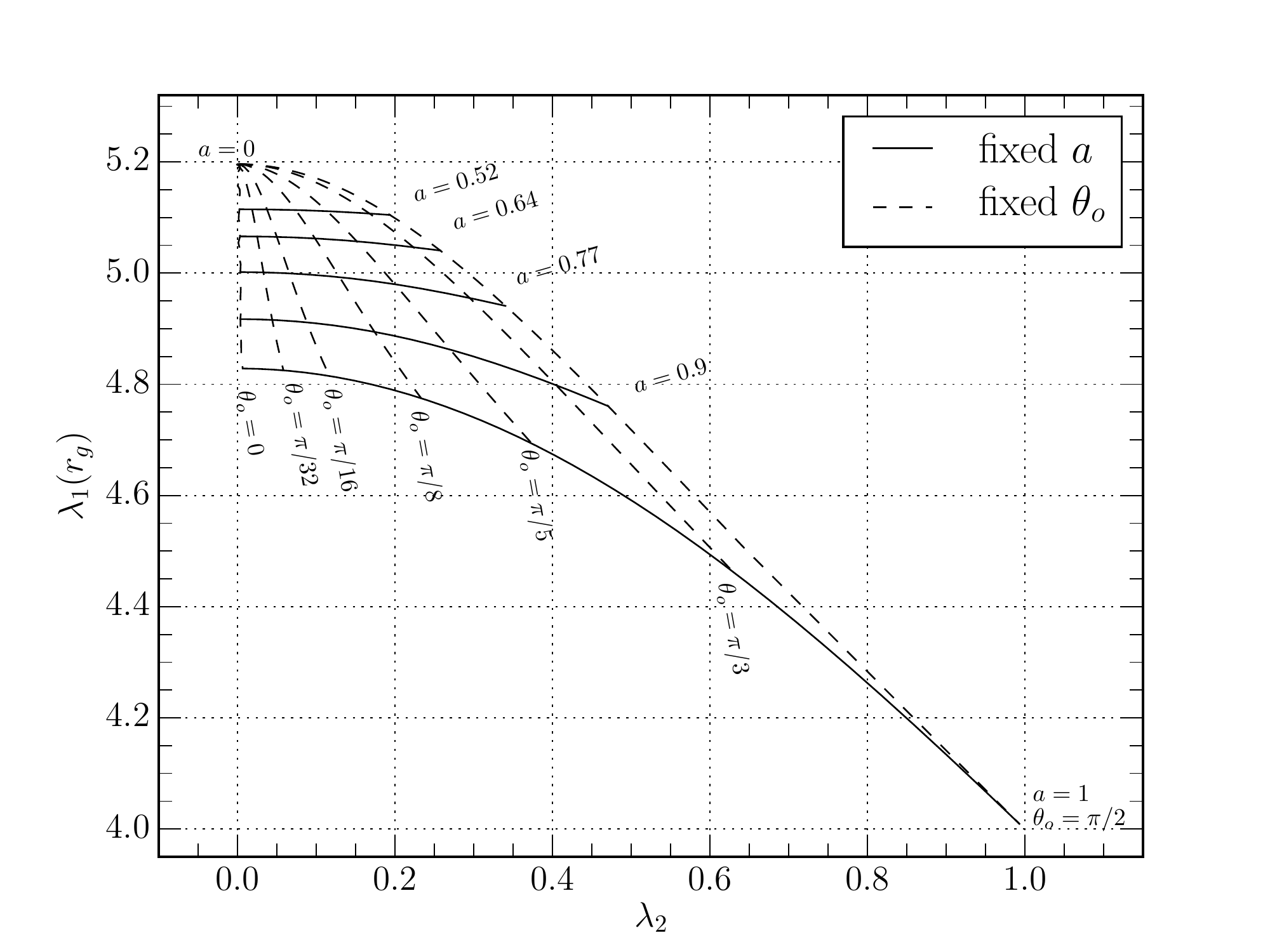}
\caption{
Best-fitting convex hull \limacon parameters $(\lambda_1, \lambda_2)$ for all spins and inclinations (a displacement parameter $\delta$ is fitted but not shown). As for the \limacon, the mapping uniquely determines the black hole spin and inclination if the mass-to-distance ratio (i.e., $r_{\rm g}$) is known. Moreover, the Kerr metric occupies only a small part of the $(\lambda_1, \lambda_2)$ parameter space, so a convex hull \limacon fit can be used to assess the validity of a Kerr assumption for the shadow and to quantify departures from Kerr. 
}
\label{fig:altgrav_convexhull}
\end{figure}

While the diagnostic quantities of the convex hull \limacon are not as trivial to derive as those of the ordinary \limacon, the convex hull \limacon similarly constrains the Kerr metric tightly in the $(\lambda_1, \lambda_2)$ parameter space, as shown in \autoref{fig:altgrav_convexhull}. 

\section{Discussion}
\label{sec:discussion}

\subsection{Comparison of Polar Approximations}

\begin{figure*}[t]
\centering
\includegraphics[width=0.325\textwidth]{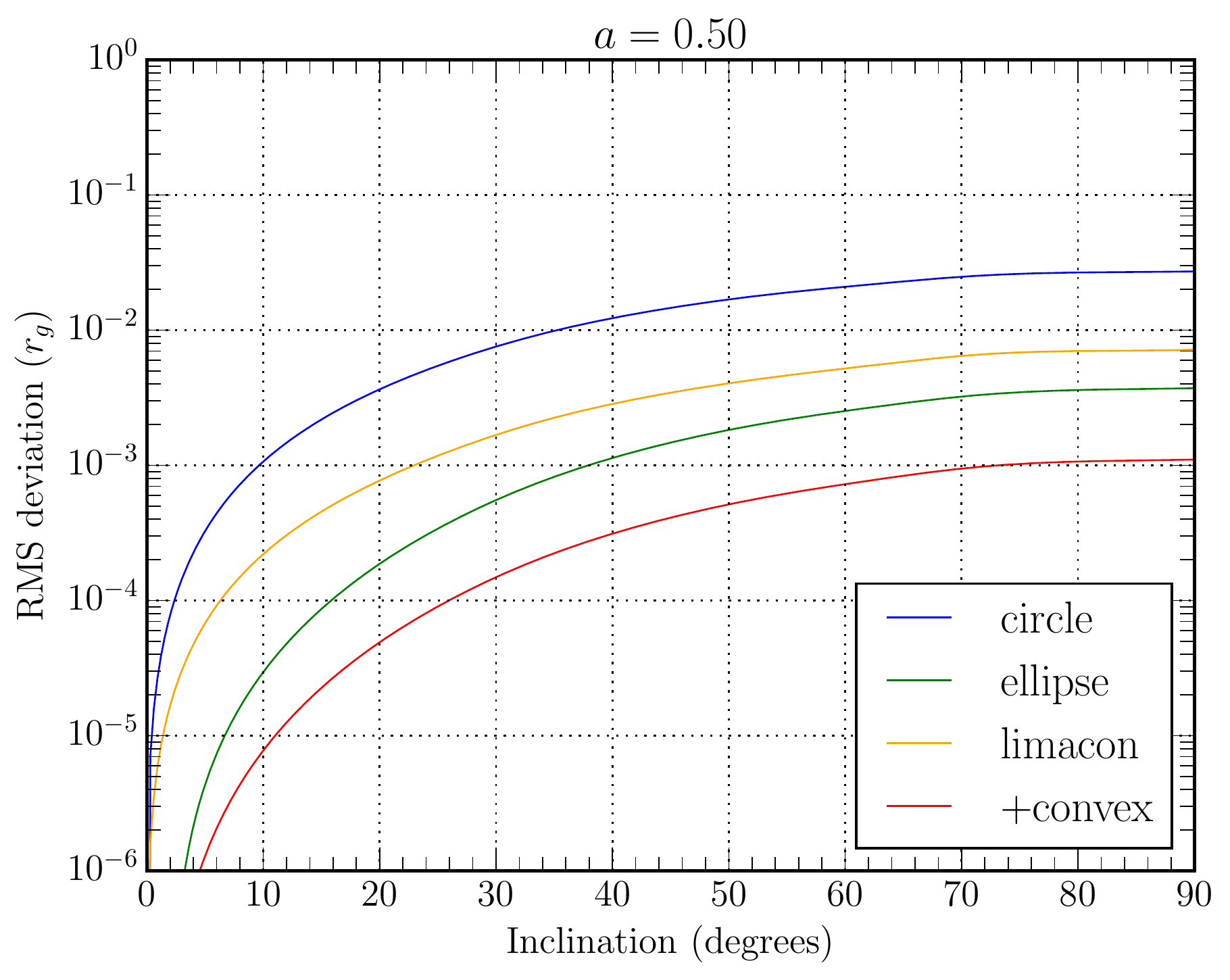}
\includegraphics[width=0.31\textwidth]{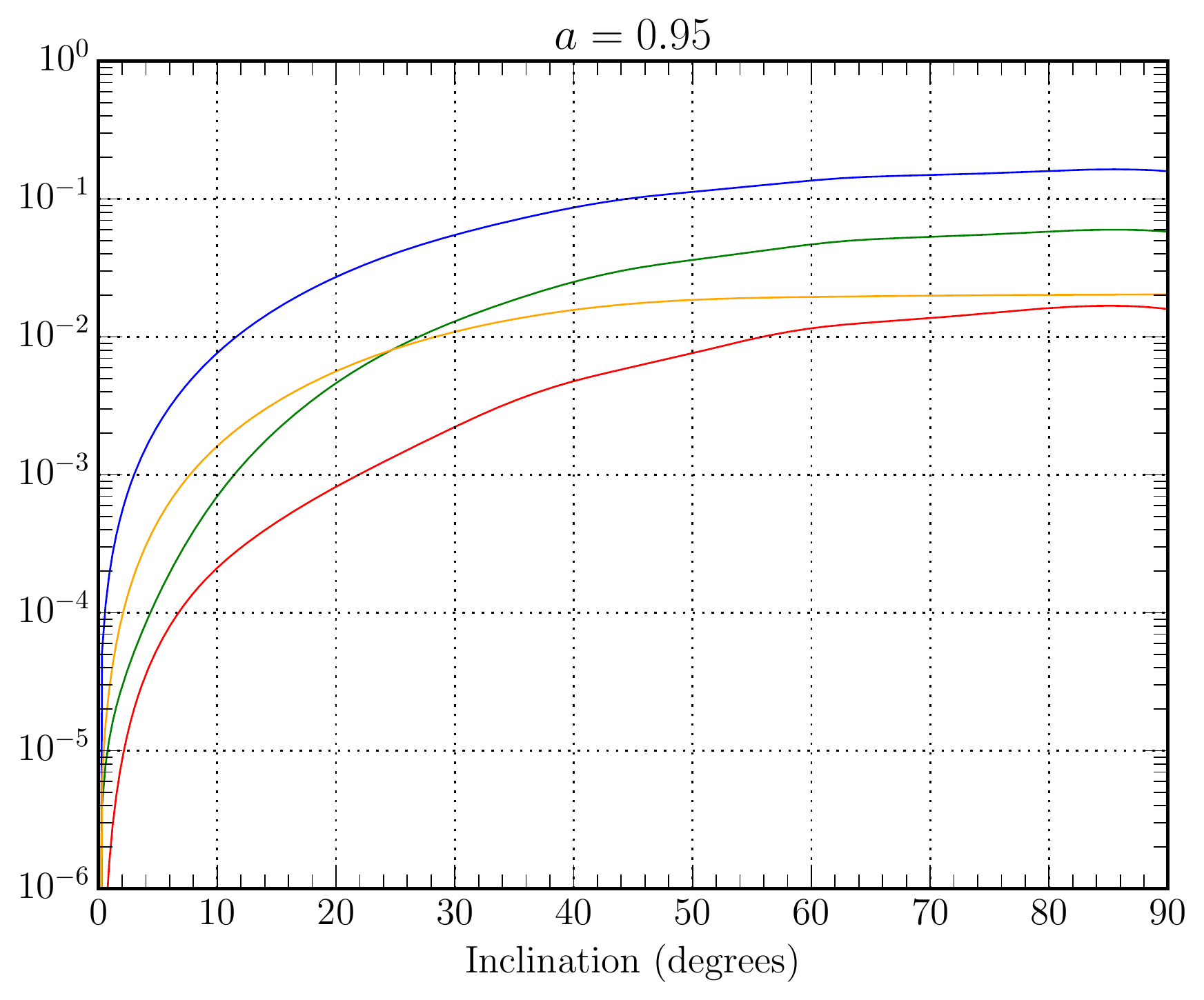}
\includegraphics[width=0.31\textwidth]{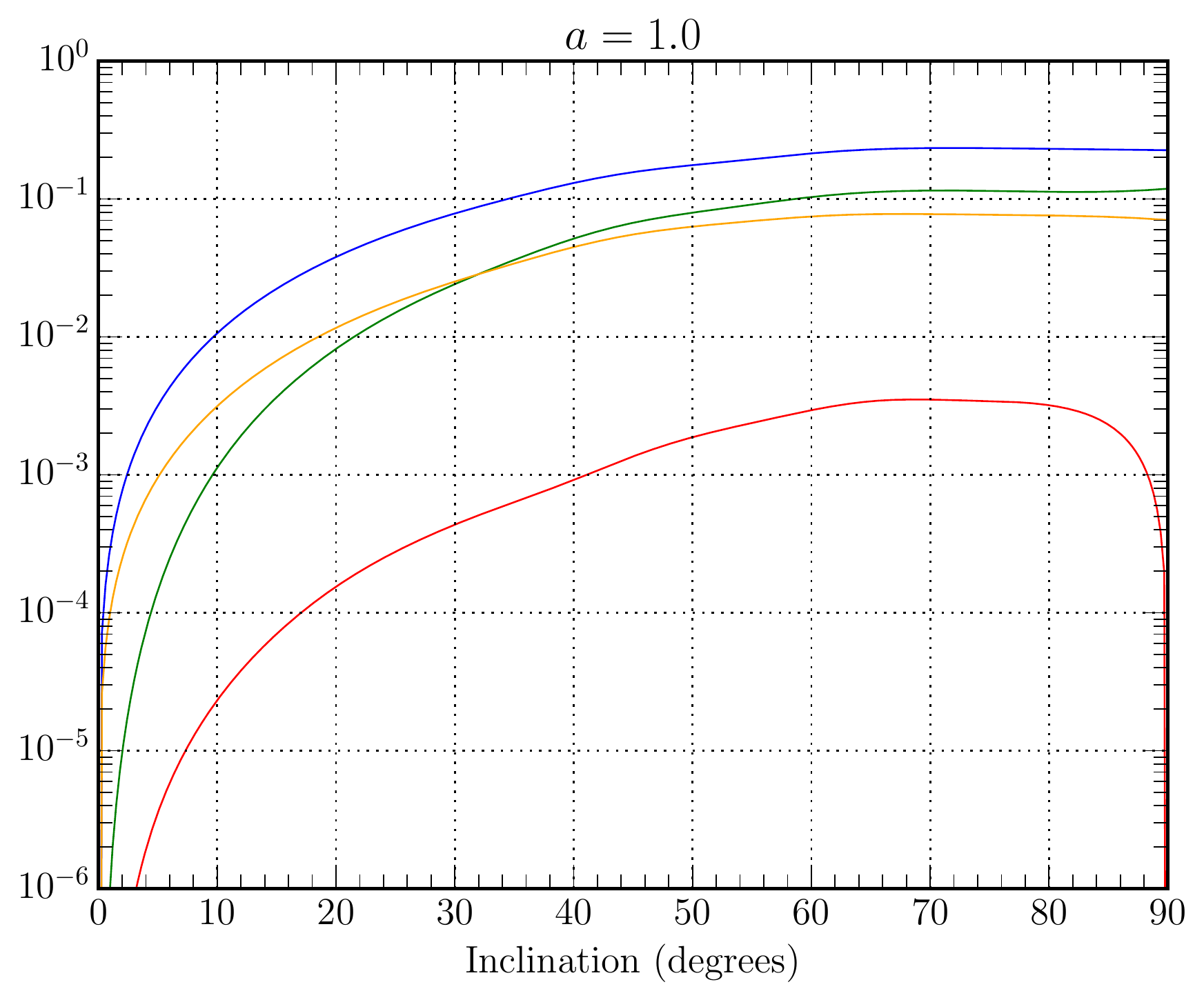}
\caption{
Accuracy of best-fitting circle (blue), ellipse (green), \limacon (orange), and convex hull \limacon (red) approximations as a function of inclination for $a=0.5$ (left), $a=0.95$ (center), and $a=1$ (right).}
\label{fig:compare_fits}
\end{figure*}

\autoref{fig:compare_fits} compares the accuracy for each of our parametric shadow approximations as a function of spin and inclination. For low spin, $a\lsim 0.5$, the black hole shadow is nearly circular and all approximations are excellent (even a circle fits reasonably well). At high spin, the circle ceases to be a reasonable fit except at very low inclination, and it has a fractional radial RMS of approximately 4\% at high inclination.  The \limacon, which can produce a flat left edge, outperforms the ellipse at high spin and inclination, while the convex hull \limacon significantly outperforms both the ellipse and \limacon at all inclinations for maximal spin.

\subsection{Implications of Accurate Polar Approximations}

The accuracy of simple curves such as the \limacon in describing the shadow has subtle implications for the general shadow geometry. For example, the \limacon is defined by only two parameters, implying that the displacement is determined entirely by the mean radius and asymmetry. This is analogous to the exact shadow, in which the shape is determined by the two parameters of black hole mass and spin. The \limacon relationship can be expressed using ``observable'' quantities such as the perpendicular radius $r_{\perp}$ and the area $A = \pi \lambda_1^2 \left( 1 + \lambda_2^2/2 \right)$:
\begin{align}
    \label{eq:Delta_A_Relationship}
    \Delta &= \sqrt{2} r_\perp \sqrt{ \frac{A}{\pi r_{\perp}^2} -1}.
\end{align}
This relationship is exact for a \limacon \change{with $\lambda_2 < 1/2$} and is approximately true for the black hole shadow ($\Delta$ is correct to within $0.4 r_{\rm g}$ and has a fractional error smaller than ${\approx}20\%$ for all values of spin and inclination). It shows that a shadow displacement only occurs when there is asymmetry and that the two effects are intimately linked. This relationship also provides a simple way to estimate the black hole's location on an image from measurements of the shadow size and shape.

\subsection{Degeneracies of Inferred Black Hole Parameters}

Another useful application of our \change{geometrical shadow approximations is to assess the degeneracies in spin and inclination. Apart from the trivial inclination degeneracy at zero spin, the shadow has a single additional discrete degeneracy for supplementary inclination angles: $\inclination \rightarrow \pi - \inclination$ \citep[see, e.g.,][]{Mars_2018}. However, if the mass-to-distance ratio of the black hole is unknown, then shadows that differ by a constant rescaling are indistinguishable (e.g., all face-on observers see a perfectly circular shadow, regardless of spin). Moreover, identification of the coordinate origin is impractical, so shadows that differ only by a displacement will likely be indistinguishable. Finally, shadows that may not be formally degenerate may still have differences that are so small as to be indistinguishable for all practical purposes. Our framework allows us to assess all these cases.}

\change{These various classes of denegeracies are evident in Figures \ref{fig:altgrav} and \ref{fig:altgrav_convexhull}. For example, 
the convex hull \limacon\ fits make no assumptions about knowledge of the coordinate centroid. For both the \limacon\ and the convex hull \limacon, a~priori knowledge of the mass-to-distance ratio is necessary to place a measurement of $\lambda_1$ in units of $r_{\rm g}$, while $\lambda_2$ is dimensionless and requires no a~priori knowledge. }

\change{For instance}, because of the a priori estimates of $r_{\rm g}$ for \sgra\ \change{using stellar orbits}, measuring the (convex hull) \limacon\ parameters $\lambda_1$ and $\lambda_2$ for \sgra\ would provide unambiguous estimates for both the spin and inclination of the black hole \change{(up to the single discrete degeneracy for supplementary inclination angles)}. Indeed, a measurement of $\lambda_1$ alone is sufficient to narrowly constrain the spin and (if $\lambda_1 < 4.83 r_{\rm g}$) to give a lower bound on the inclination. 

Likewise, while the \change{supermassive black hole} in M87 does not have a strong prior on $r_{\rm g}$ and, thus, only $\lambda_2$ can be measured directly, it does have a tightly constrained jet inclination. A measurement of $\lambda_2$ would then determine the spin magnitude $a$ under the assumption that the spin axis is aligned with the jet. 

For sources with no prior constraints on $r_{\rm g}$ or $\inclination$, a measurement of $\lambda_2$ provides a lower limit on both spin and inclination but does not absolutely determine either. Larger values of spin and inclination would correspond to smaller values of $\lambda_1/r_{\rm g}$ and, thus, larger values of the black hole mass. Because the convex hull \limacon is an excellent approximation to the true shadow, these degeneracies are unlikely to be resolved even with highly accurate measurements of the shadow shape, although they could be broken with complementary measurements (e.g., temporal signatures, modeling of the complete system, or relationships between sequential photon subrings).

\section{Summary}
\label{sec:summary}

We have shown that simple polar curves can provide excellent approximations to the exact boundary of the shadow from a Kerr black hole (\autoref{eq:alpha_beta}). A suitable polar curve must account for three geometric properties of the black hole shadow: diameter, asymmetry, and horizontal displacement. For a Kerr black hole, the shadow diameter and displacement are directly proportional to the black hole mass, while all three properties are affected by the spin magnitude and inclination.

We first explored fitting the shadow with a displaced ellipse, which provides a marked improvement in fit quality over a circle (see Figure~\ref{fig:compare_fits}) and offers convenient analytic expressions for directly fitting interferometric quantities. The RMS error in radius, weighted by arc length, is less than $0.1 r_{\rm g}$ for all spins and inclinations, corresponding to a fractional error of only 2\%. For a source viewed at low inclination, as is expected for the supermassive black hole in M87, the fractional RMS error is less than $0.1\%$ (corresponding to $ 0.02\,\mu{\rm as}$ for M87). Hence, we expect the ellipse to provide an excellent shadow approximation when fitting parametric models to EHT observations of M87, with sufficient complexity to provide estimates of the shadow asymmetry and, hence, the black hole inclination and spin. A set of ellipse parameters uniquely determines the black hole spin if the \change{mass-to-distance ratio} or inclination is known a priori (as is relevant for \sgra\ and M87, respectively).

Next, we explored fitting the shadow with a family of curves known as \limacons, which are two-parameter polar curves, \change{as originally proposed and studied by \citet{vries_2003}}. We showed that a \limacon often fits the shadow better than an ellipse despite requiring one fewer parameter. In addition, \limacons have an inherent relationship between displacement and asymmetry just as Kerr black holes do. Hence, they provide convenient expressions for this relationship for the exact black hole shadow (e.g., \autoref{eq:Delta_A_Relationship}). 

Finally, we explored fitting the shadow with a generalized \limacon, adding a displacement parameter and taking a convex hull. Especially at high spin, this convex hull \limacon provides a significantly better fit than the unmodified \limacon or ellipse. The best-fitting convex hull \limacon has a fractional RMS radial residual of less than $0.3\%$ for all black hole spins and inclinations. It also gives an exact description of the shadow in the limit of a maximally spinning black hole viewed edge-on. 

These polar curves offer many utilities including parametric model fitting to interferometric data, simple diagnostics of the shadow properties, and a low-dimensional geometrical representation of the full shadow shape (just as the exact shadow is defined solely by the mass and spin of the black hole). They also reveal degeneracies that are expected when inferring a black hole's parameters from images of its shadow. For instance, because a displacement cannot be directly observed and the convex hull \limacon provides an excellent fit to the shadow shape, an accurately measured shadow only gives a bounded range for the black hole mass as well as lower limits on the spin and inclination (see \autoref{fig:altgrav_convexhull}). However, an a priori measurement of the black hole mass, spin, or inclination is sufficient to break this degeneracy, yielding the other two parameters from the measured shadow. Finally, our curves provide a flexible parametric framework to assess whether a particular shadow is compatible with the Kerr metric. While simple circular models were suitable for the initial EHT observations of M87, these curves offer a pathway to interpret the increasingly sensitive observations expected in the coming decades, with correspondingly more stringent demands for model fidelity.

\acknowledgements{
We acknowledge Jonathan Delgado, Chi-Kwan Chan, Paul Tiede, and Roman Gold for helpful discussions. This work was supported by the NSF (AST-1716536, AST-1440254, AST-1935980, and OISE-1743747) and the Gordon and Betty Moore Foundation (GBMF-5278). 
This work was carried out at the Black Hole Initiative, Harvard University, which is funded by grants from the John Templeton Foundation and the Gordon and Betty Moore Foundation to Harvard University.
}

\bibliographystyle{yahapj}
\bibliography{ref}

\end{document}